\documentclass[12pt]{iopart}
\usepackage{amsfonts} 
\usepackage{graphicx} 
\usepackage{graphics}       
\usepackage[colorlinks,linkcolor=blue,citecolor=magenta]{hyperref} 
\usepackage{subfigure}
\usepackage{iopams}  
\newcommand{\p}{\mathrm{p}}
\begin{document}

\title[Revealing a quantum feature of dimensionless uncertainty...]{Revealing a quantum feature of dimensionless uncertainty in linear and quadratic potentials by changing potential intervals}

\author{R. Kheiri}

\address{Department of physics, Isfahan University of Technology, Isfahan 84156-83111, Iran}
\ead{r.kheiry@ph.iut.ac.ir}
\vspace{10pt}
\begin{indented}
\item[]June 2016
\end{indented}

\begin{abstract}
As an undergraduate exercise, in an article [Am. J. Phys. {\bf 80} (2012), 708--14], quantum and classical uncertainties for dimensionless variables of position and momentum were evaluated in three potentials: infinite well, bouncing ball, and harmonic oscillator. While original quantum uncertainty products depend on $\hbar$ and the number of states ($n$), dimensionless approach makes comparison between quantum uncertainty and classical dispersion possible by excluding $\hbar$. But the question is whether the uncertainty still remains dependent on quantum number $n$. In the above mentioned article, there lies this contrast; on the one hand, dimensionless quantum uncertainty of potential box approaches classical dispersion only in the limit of large quantum numbers ($n\to\infty$) -- consistent with the correspondence principle. On the other hand, similar evaluations for bouncing ball and harmonic oscillator potentials are equal with their classical counterparts independent of $n$. This equality may hide the quantum feature of low energy levels. In the current study, we've changed the potential intervals in order make them symmetric for the linear potential and non-symmetric for the quadratic potential. As a result, it is shown in this paper that the dimensionless quantum uncertainty of these potentials in the new potential intervals is expressed in terms of quantum number $n$. In other words, the uncertainty requires the correspondence principle in order to approach the classical limit. Therefore, it can be concluded that the dimensionless analysis as a useful pedagogical method does not take away the quantum feature of n-dependence of quantum uncertainty in general. Moreover, our numerical calculations include the higher powers of position for potential.
\end{abstract}

%
\vspace{2pc}
\maketitle
%
\section{Introduction}
Since the advent of quantum mechanics, there have been substantial works attempting to approximate the novel quantum notions to the old classical concepts. Among them, we can refer to correspondence principle \cite{bohr}, bold classical path \cite{ehrenfest}, quantum Virial theorem \cite{Fock}, joint distribution function of position and momentum \cite{Wigner}, Dirac's analogy \cite{Dirac}, etc. Such theoretical researches are still referred to by scholars as major sources on the foundations of quantum mechanics, among which we can mention studies \cite{Ballentine,Mello,Bamber}.

On the other hand, some studies have attempted to build a probabilistic structure for inherently predetermined variables of classical mechanics in order to compare essentially statistical quantum observables. According to Curtis and Ellis \cite{Curtis}, an introductory physics course utilizes a historical Newtonian approach which is identified by instantaneous values for position, speed and acceleration, whereas a quantum course is featured by conceptually probabilistic observables.
\begin{quote}
\emph{``Although the traditional tracking of instantaneous positions (like a series of snapshots)
has conceptual advantages, the use of position probabilities (like a time exposure that reveals
motion through the degree of overexposure) has other advantages. It allows macroscopic and
microscopic objects to be studied by similar techniques. It is easily extended to include
many-body interactions, since the probability distributions can be superimposed. ... Another advantage of the probabilistic
approach is that it provides a convenient way of performing numerical computations for
potentials that do not have an analytic solution.''}
\end{quote}
This approach leads to the comparison of classical and quantum probabilities in high energies by applying the correspondence principle. In other words, quantum distribution wiggles so rapidly in a scale set by the classical amplitude that only its mean can be detected at these scales, and this agrees with classical probability \cite{Shankar}. Hence, squared wave function and classical probability density, if averaged in a finite interval, seem to be indistinguishable for large quantum number $n$ of one-dimensional problems. In a study conducted by Robinett \cite{Robinett}, this point of view was demonstrated by several examples.

Here we highlight the structure of classical probability in more details. Basically, position probability density of one point particle (or an ensemble of identical particles) with its motion equation in one dimension $x=\mathcal{X}(t)$ can be written as
\begin{equation}
p(x,t)\propto \delta (x-\mathcal{X}(t)).
\label{eqpcomplex}
\end{equation}
On the other hand, for a time-independent potential, we have $p(x,t) = p(x) T(t)$ and therefore, we can track time-independent position probability density $p(x)$ via the probability of finding the particle in a small region, $dx$, proportional to the amount of time, $dt$, the particle elapse there.
\[ {\rm{probability}} [(x,x+\rmd x)]\equiv p(x) \rmd x ,\]
\begin{equation}
p(x) \rmd x \propto \rmd t \Rightarrow p(x) \propto \frac{1}{v(x)} \propto \frac{1}{\sqrt{E -U(x)}},
\label{vequation}
\end{equation}
, where $v$ is speed in terms of $x$, and $E$ and $U$ stand for total and potential energy, respectively.
The equivalency of equations \eref{eqpcomplex} and \eref{vequation} concerning position can be reached by the property of delta function, as can be seen in the following equation: 
\begin{equation}
\delta (x-\mathcal{X}(t)) = \sum_{i} \frac{1}{\left| \mathcal{X}' (t_{i})\right|} \: \delta (t- t_{i})\qquad , \qquad t_{i} = \mathcal{X}^{-1}(x).
\end{equation}
For time-independent potentials, we have a single-valued function for speed $v(x)$ in terms of $x$. Therefore
\begin{equation}
v(x) \; \delta (x-\mathcal{X}(t)) = \sum_{i} \delta (t- t_{i}) \quad , \quad v(x) =\left| \mathcal{X}' (t_{i})\right| .
\label{delta2}
\end{equation}
Now, substituting \eref{eqpcomplex} on the left side of \eref{delta2} and integrating in respect to $dt$ on both sides of \eref{delta2} result in \eref{vequation}.
As an example, for a Harmonic oscillator $\mathcal{X}(t)=A \sin (\omega t \pm \phi )$ and $v(x) = A \omega \cos ({\sin}^{-1} \frac{x}{A}) = \omega \sqrt{A^2 - x^2 }$, hence $t_{i} = \frac{1}{\omega} [\sin^{-1}\frac{x}{A}\mp \phi ] $ is unique so that we do not have summation on the right side of \eref{delta2}, unlike the free fall equation of motion, which leads to two terms for this summation.

We can also derive similar equations for the classical probability density of momentum. For instance, an analogous equation for \eref{vequation} in the momentum space would be \cite{Robinett}
\begin{equation}
p(\p) \propto \frac{1}{|F(x)|}
\end{equation}
, where $F$ refers to the force. 

\Eref{delta2} can be interpreted as a classical formula where determined position and speed for point particle are matched together. However, knowing this determination can not prevent us from calculating classical dispersion as ${\left( \Delta {x} \right)}_{\rm cl}^{2} \: {\left( \Delta {\p} \right)}_{\rm cl}^{2}$ by means of averaging in respect to classical probability densities, so that it can be compared with quantum uncertainty relation ${\left( \Delta \hat{x} \right)}_{\rm qm}^{2} \: {\left( \Delta \hat{\p} \right)}_{\rm qm}^{2}$.

As Hamiltonian approach can substituted for Newtonian mechanics, it can also be applied for the classical probability of the phase space in case we write a delta function on the subject of Hamiltonian $\mathcal{H} = \frac{{\p}^2}{2m} + U(x) $ in a given Energy
\begin{equation}
p(x, \p) \propto \delta \left[ \frac{{\p}^2}{2m} + U(x) - E \right]
\end{equation}
, where constant $E$ in a given potential, directly refers to one particle or an ensemble of identical particles.

Straightforwardly, classical averaging can be obtained just in terms of position probability density \cite{Devi} (see \ref{averagefequation}) 
\begin{eqnarray}
\fl \left\langle f(x,\p )\right\rangle = \frac{1}{2} \int \rmd x\, p(x) \left[ f \left( x , - \sqrt{2m[E -U(x)]} \right) + f \left( x , \sqrt{2m[E -U(x)]} \right)  \right] 
 \label{eqaverage}
\end{eqnarray}
Devi and Karthik \cite{Devi} have chosen dimensionless variables of position and momentum as
\begin{equation}
X=x/{x}_{max}$, $P=\p/{\p}_{max},
\label{eqdimensionless}
\end{equation}
, where $x_{max}$ and ${\p}_{max}$ are classical return points;
then they analytically compared classical dispersion ${\left( \Delta {X} \right)}_{\rm cl}^{2} \: {\left( \Delta {P} \right)}_{\rm cl}^{2}$
with quantum uncertainty of ${\left( \Delta \hat{X} \right)}_{\rm qm}^{2} \: {\left( \Delta \hat{P} \right)}_{\rm qm}^{2}$ for three potentials of infinite well, bouncing ball, and harmonic oscillator.

The following dimensionless classical position and momentum dispersions
\begin{equation}
\eqalign{{\left( \Delta {X} \right)}_{\rm cl}^{2} = {\left\langle X^{2} \right\rangle}_{\rm cl} - {\left\langle X \right\rangle}_{\rm cl}^{2}, \cr 
{\left( \Delta {P} \right)}_{\rm cl}^{2}= {\left\langle P^{2} \right\rangle}_{\rm cl} - {\left\langle P \right\rangle}_{\rm cl}^2 ,}
\end{equation}
could be evaluated using \eref{eqaverage}, whereas dimensionless quantum uncertainty could be evaluated by averaging on squared wave functions, so $x_{max}(n)$ and ${\p}_{max}(n)$ are derived as functions of $n$ from $E_{qm} = E_{cl}$.

With some generalization, their results for three mentioned potentials are as follows:
\begin{itemize}
\item Symmetric or non-symmetric potential box, where $A$ is a constant
\begin{equation}
\fl \eqalign{U(x) = \cases{0&for $-A\leq x \leq A$ or $0 \leq x \leq A$\\
\infty & otherwise \\}, \cr
\lim_{n\to\infty} {\left( \Delta \hat{X} \right)}_{\rm qm}^{2} \: {\left( \Delta \hat{P} \right)}_{\rm qm}^{2} = {\left( \Delta {X} \right)}_{\rm cl}^{2} \: {\left( \Delta {P} \right)}_{\rm cl}^{2}=\cases{1/3 & symmetric\\
1/12 & non-symmetric \\}.}
\label{eqbox}
\end{equation}
\item Bouncing ball\footnote{The term ``bouncing ball potential'' describes a uniform gravitational field above a rigid flat surface if $ax\rightarrow mgz$ in \eref{eqbouncing}. Quantum solutions for bouncing ball potential are just the odd states of the symmetric linear potential. For more information about this potential, see \cite{Robinett, Gea, Goodmanson}. } (non-symmetric linear) potential
\begin{equation}
\eqalign{U(x) = \cases{\infty &for $x<0$\\
ax &for $x\geq 0$\\}, \cr
{\left( \Delta \hat{X} \right)}_{\rm qm}^{2} \: {\left( \Delta \hat{P} \right)}_{\rm qm}^{2} = {\left( \Delta {X} \right)}_{\rm cl}^{2} \: {\left( \Delta {P} \right)}_{\rm cl}^{2} = \frac{4}{135}.}
\label{eqbouncing}
\end{equation}
\item Harmonic oscillator (symmetric quadratic) potential
\begin{equation}
\eqalign{U(x) =a {x^2}, \cr
{\left( \Delta \hat{X} \right)}_{\rm qm}^{2} \: {\left( \Delta \hat{P} \right)}_{\rm qm}^{2} = {\left( \Delta {X} \right)}_{\rm cl}^{2} \: {\left( \Delta {P} \right)}_{\rm cl}^{2} = \frac{1}{4}}
\label{eqharmonic}
\end{equation}
\end{itemize}
, where $a$ is a positive integer constant which is neutral when choosing dimensionless variables (just like $m$ and $\hbar $ in the relevant Schrodinger equation of \ref{appedixmatrix}).

Evaluations of potential box \eref{eqbox} approach classical dispersion only in the limit of
large $n$, which is consistent with the correspondence principle. Namely, its dimensionless quantum uncertainty depends on quantum number $n$. But for bouncing ball \eref{eqbouncing} and harmonic oscillator \eref{eqharmonic} potentials, quantum uncertainties are independent of $n$ and exactly equal to classical evaluations. The reason why the quantum uncertainty of bounding ball and harmonic oscillator potentials are to be independent of $n$ whereas that for the potential box is not (or further, for other powers of $x$ in the potential except for powers 1 and 2) could be left as an exercise. [hint: use $<\hat{x}> \propto <\hat{U}>$ or $<{\hat{x}}^{2}> \propto <\hat{U}>$ or $<{\hat{x}}^{2}> \propto <{\hat{U}}^{2}>$ and consider Virial theorem].   

We could obtain uncertainty relations for original variables of $\hat{x}$ and $\hat{\p}$ instead of dimensionless equations \eref{eqbox}, \eref{eqbouncing}, and \eref{eqharmonic}. Original uncertainty products
\begin{equation}
{\left( \Delta \hat{x} \right)}_{\rm qm}^{2} \: {\left( \Delta \hat{\p} \right)}_{\rm qm}^{2} \geq \frac{{\hbar}^2}{4}
\label{uctyeq}
\end{equation}
depend on $\hbar$; also it depends on $n$ since we have integrands as $\psi_n (x)$ and we take integrals in respect to variable $x$; then $n$ is preserved. For example, we can show that ${\left( \Delta \hat{x} \right)}_{\rm qm}^{2} \: {\left( \Delta \hat{\p} \right)}_{\rm qm}^{2} =(n+ 1/2)^{2} {\hbar}^2 $ for the symmetric harmonic oscillator \cite{Sakurai} [hint: use properties of Hermite functions or equivalently use the effect of annihilation and creation operators on a given state]. When we chose dimensionless variables of \eref{eqdimensionless} at the left side of \eref{uctyeq}, it causes the dimensional constant of $\hbar$ (or $h$) to be prevented at the right. Although excluding $\hbar$ is an advantage of the dimensionless approach of uncertainty and makes its comparison with classical dispersion possible, regarding n-dependence as an other quantum feature of quantum uncertainty, there is still the question whether this dimensionless uncertainty remains dependent on quantum number $n$ under other circumstances.

In the present study, we bring into consideration dimensionless quantum uncertainty versus dimensionless classical dispersion, as obtained by averaging on classical probability density. First of all, we numerically calculate dimensionless classical and quantum uncertainty for symmetric and non-symmetric potentials from powers of coordinate. Then, we narrowly focus on linear and quadratic potential solutions and compare them with the results of the \cite{Devi} (equations \eref{eqbouncing} and \eref{eqharmonic}). The key question is whether the dependence on $n$ for the dimensionless uncertainty of these two potentials can be retrieved if their potential intervals be changed. To seek an answer to this question in \sref{sectionlinear} of this article, we will focus on the uncertainty of symmetric linear potential and compare it with that of the non-symmetric linear potential (bouncing ball) of \eref{eqbouncing}. Also, in \sref{sectionharmonic}, we deal with the uncertainty of the non-symmetric quadratic potential and compare it with that of the symmetric quadratic potential (harmonic oscillator) of \eref{eqharmonic}. In the above mentioned sections, regarding linear and quadratic potentials, it is shown that although there is a potential interval in which dimensionless quantum uncertainty is equal with its classical counterpart --independent of quantum number $n$, according to \cite{Devi}, another interval can be chosen to reveal the dependence on $n$. Therefore, this change can be considered as an emphasis on dimensionless approach as a practical pedagogical method that does not take away the quantum feature of n-dependence in general. Moreover, the solutions of potentials with higher powers of coordinate confirm n-dependence behavior in dimensionless uncertainty. Last but not least, we discuss shortly the quantum uncertainty concept being preserved in the classical limits and also, look at the effect of measurement on the interpretation of the classical limit of a non-degenerate wave function as a classical ensemble. This study aims to emphasize on both numerical and analytical curricula. Accordingly, the first and third appendices contain analytic contexts used in the main text and the second appendix includes some basic materials used for numerical solutions of Schrodinger equation in the matrix representation method.

\section{Numerical calculations of classical and quantum uncertainties}\label{sectionnumerical}
\begin{figure}[h!]
\centering
\subfigure[]{
\includegraphics[width=7.1cm, height=5.5cm]{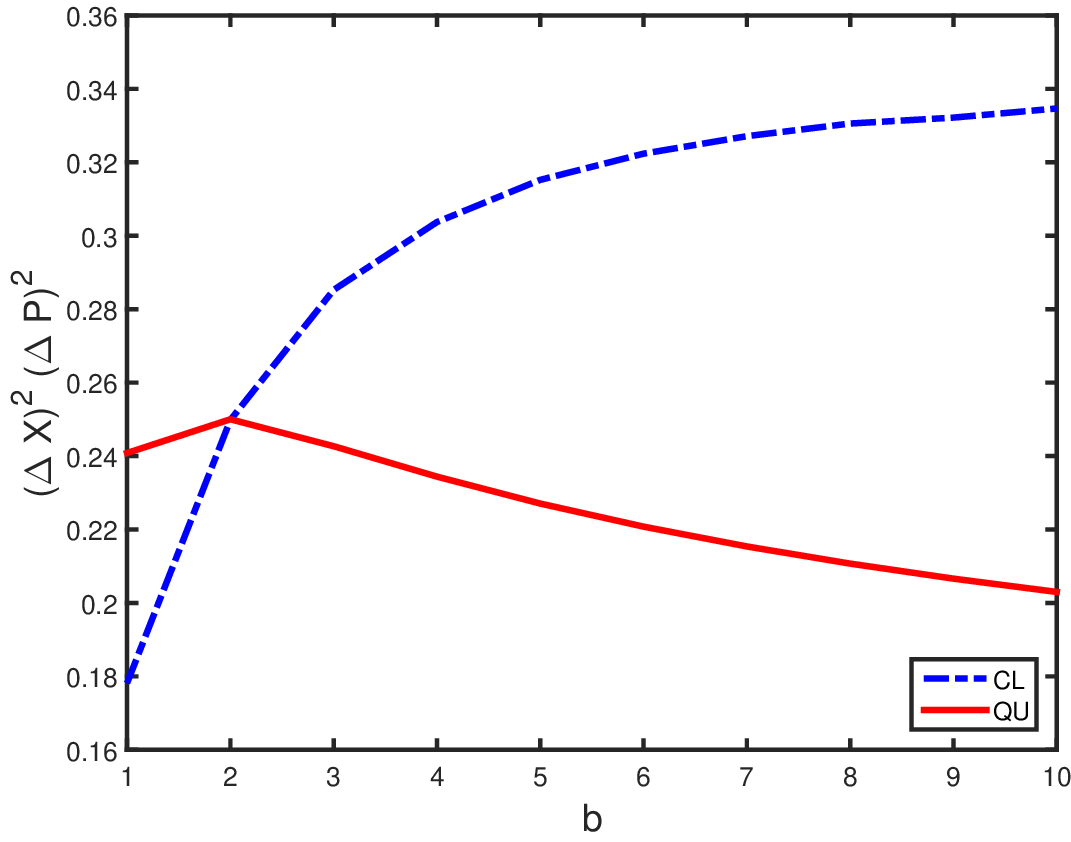}
\label{n_027000}
}
\subfigure[]{
\includegraphics[width=7.1cm, height=5.5cm]{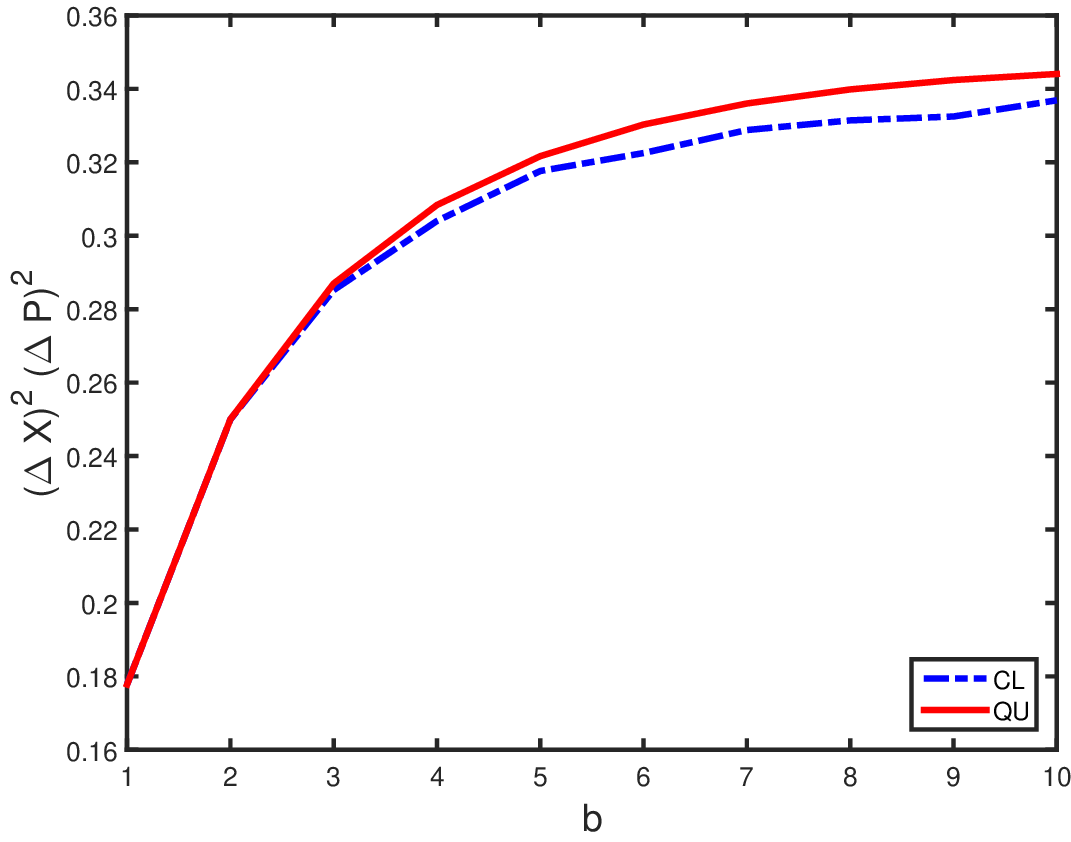}
\label{n_127000}
}
\subfigure[]{
\includegraphics[width=7.1cm, height=5.5cm]{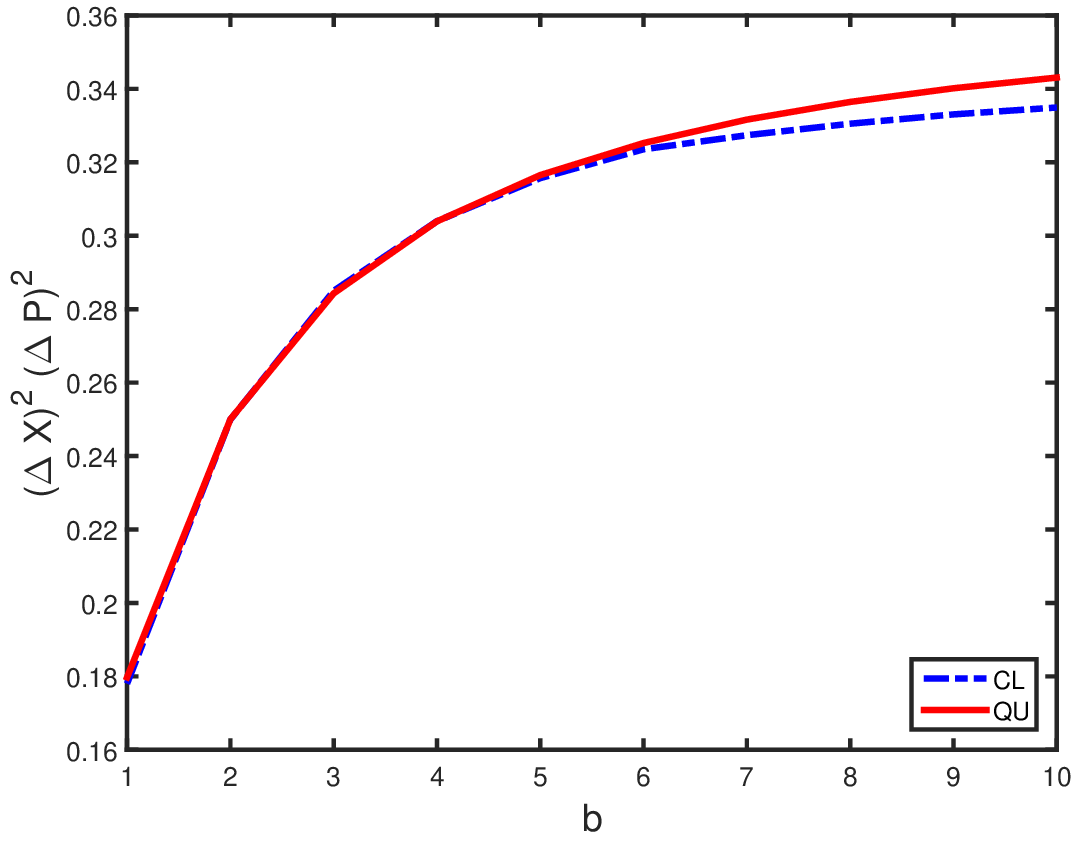}
\label{n_227000}
}
\subfigure[]{
\includegraphics[width=7.1cm, height=5.5cm]{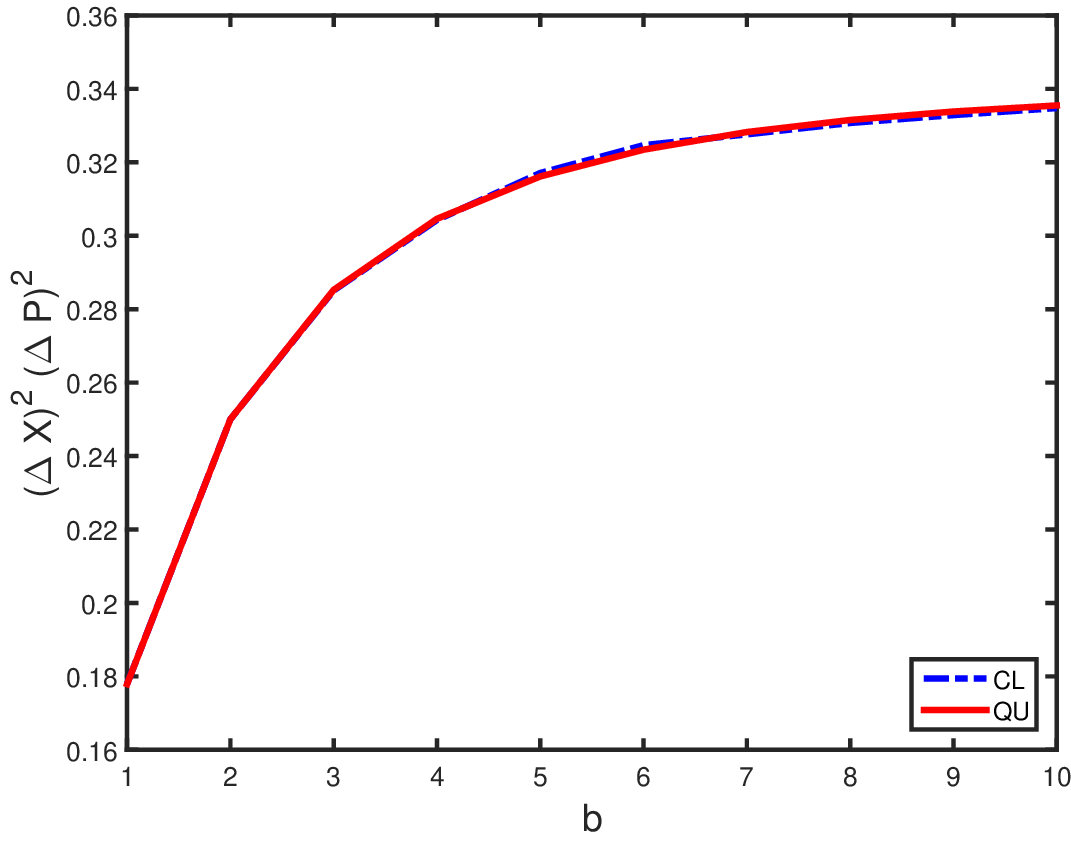}
\label{n_1027000}
}
\caption{Classical (dashed) and quantum (solid) uncertainty of dimensionless variables for symmetric potentials of equation \eref{eq4055} (horizontal axis stands for powers of $|x|$). \ref{n_027000} n=0, \ref{n_127000} n=1, \ref{n_227000} n=2, \ref{n_1027000} n=10.}
\label{gasbulbdata}
\end{figure}

\begin{figure}[h!]
\centering
\subfigure[]{
\includegraphics[width=7.1cm, height=5.5cm]{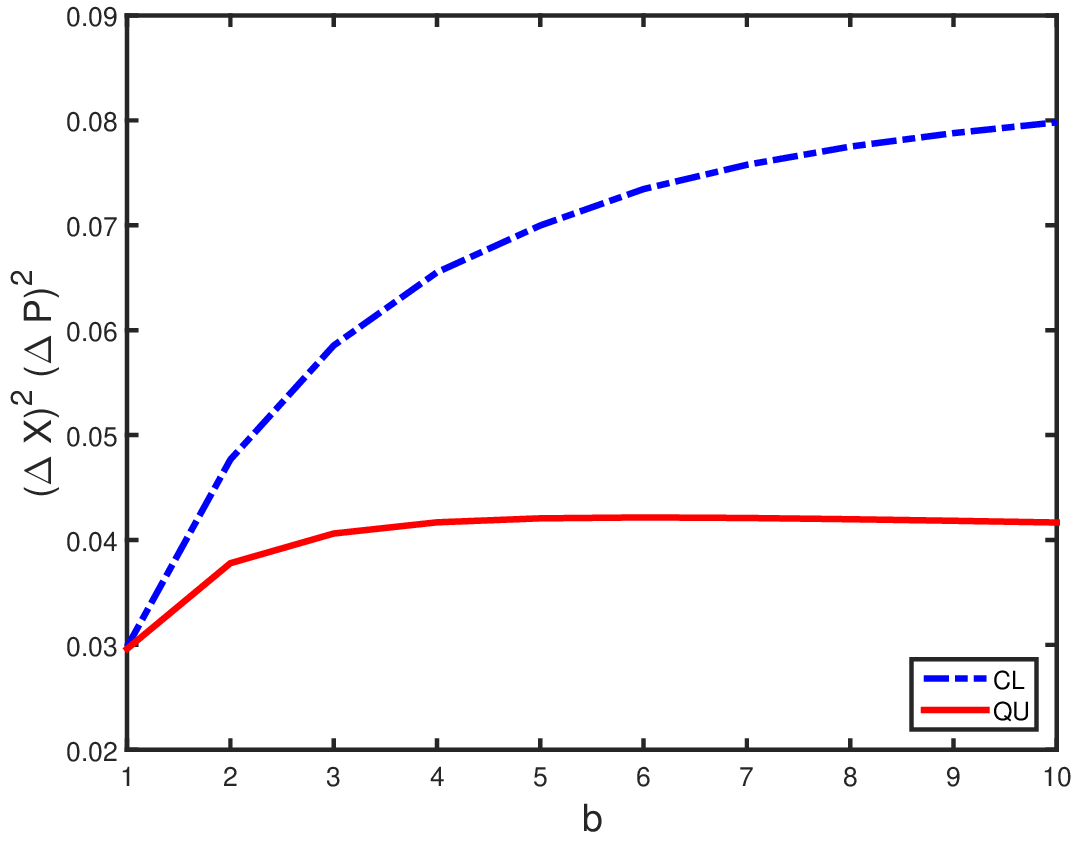}
\label{nonsymetric_n_0}
}
\subfigure[]{
\includegraphics[width=7.1cm, height=5.5cm]{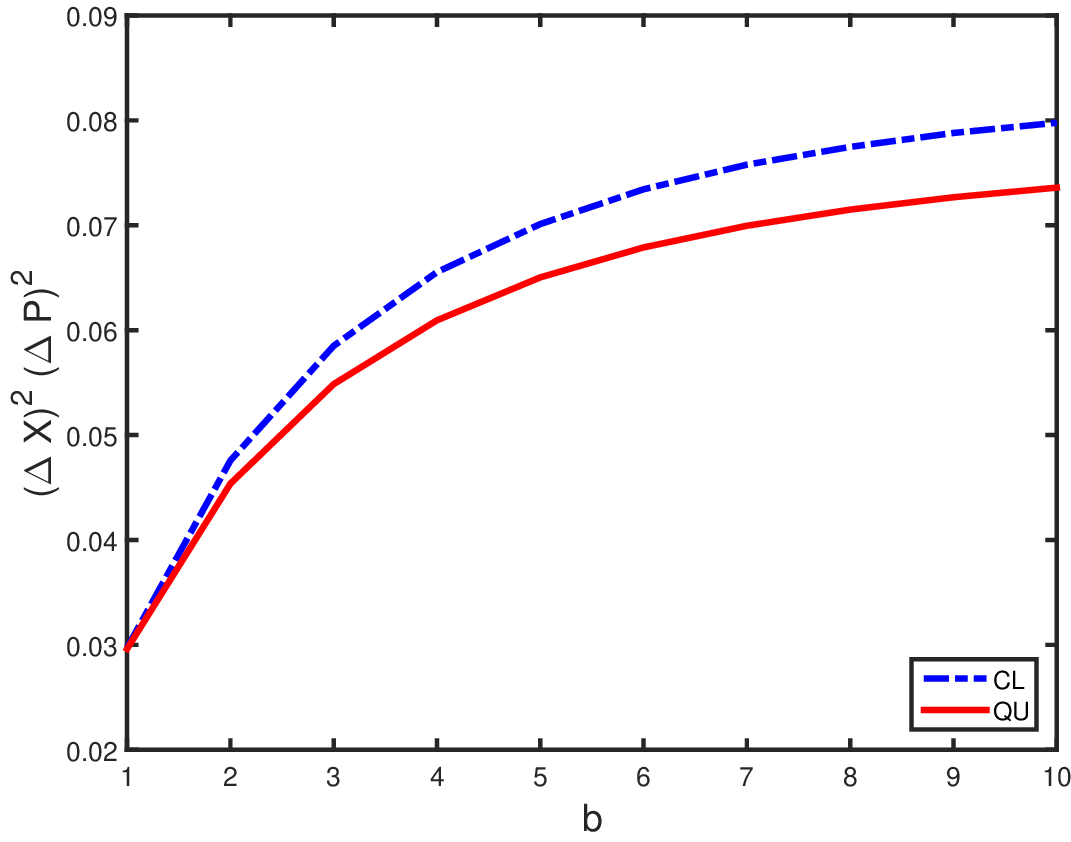}
\label{nonsymetric_n_1}
}
\subfigure[]{
\includegraphics[width=7.1cm, height=5.5cm]{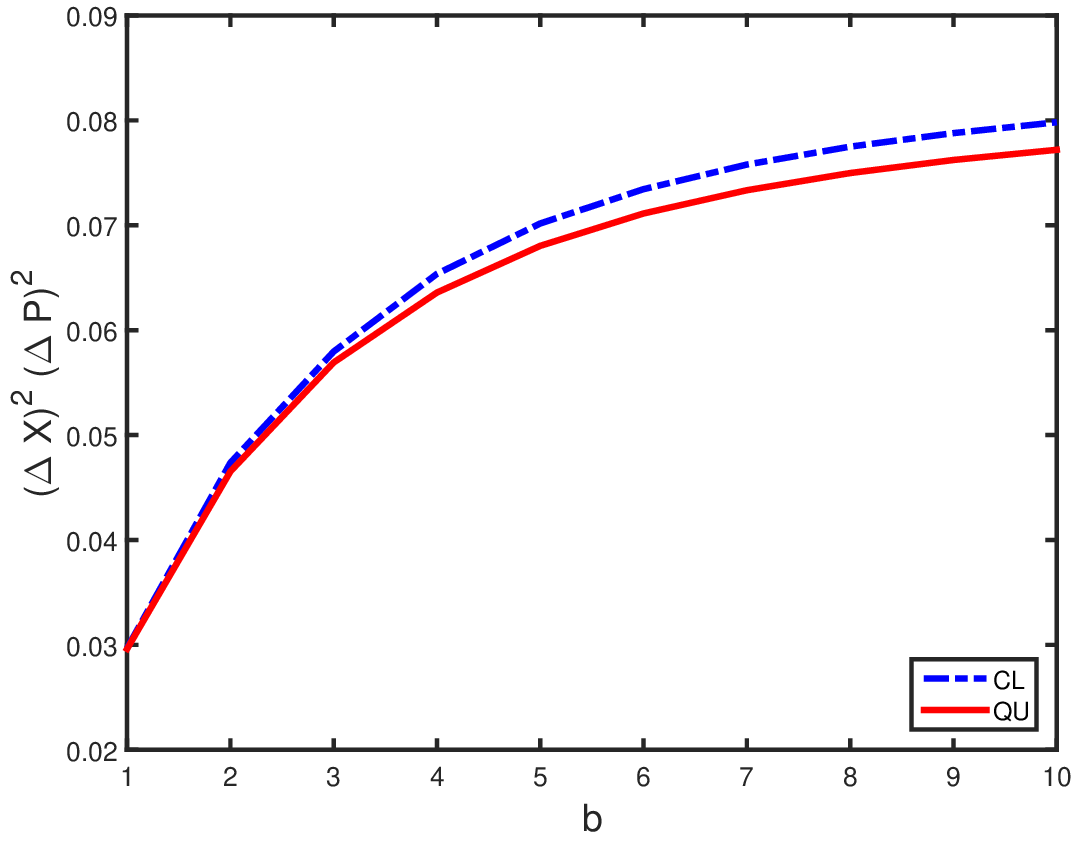}
\label{nonsymetric_n_2}
}
\subfigure[]{
\includegraphics[width=7.1cm, height=5.5cm]{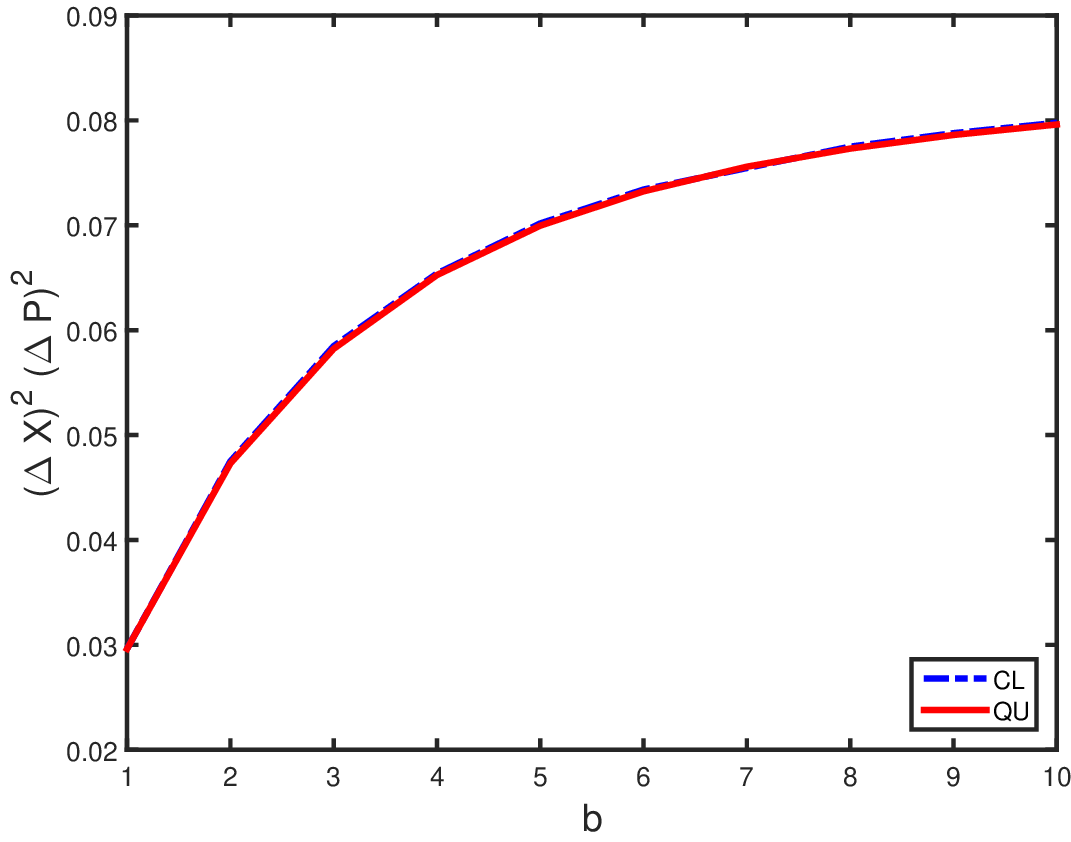}
\label{nonsymetric_n_10}
}
\caption{Classical (dashed) and quantum (solid) uncertainty of dimensionless variables for non-symmetric potentials of equation \eref{eq4056nonsymetric} (horizontal axis stands for powers of $x$). \ref{nonsymetric_n_0} n=0, \ref{nonsymetric_n_1} n=1, \ref{nonsymetric_n_2} n=2, \ref{nonsymetric_n_10} n=10.}
\label{sunsets}
\end{figure}
We consider symmetric potentials
\begin{equation}
U(x) = a {|x|}^b , \quad \textrm{where}\  b=1,2, \ldots ,10
\label{eq4055}
\end{equation}
and non-symmetric potentials
\begin{equation}
U(x) = \cases{\infty &for $x<0$ , \\
a x^b &for $x\geq 0$ \\}
\label{eq4056nonsymetric}
\end{equation}
, respectively, in the Schrodinger equation, 
\begin{equation}
- \frac{{\hbar}^2}{2m} \frac{\rmd ^{2} \psi (x)}{{\rmd x}^2} + \, U(x) \psi (x) = E \psi (x)
\label{schrodinger}
\end{equation}
and numerically calculate the amount of quantum uncertainty and classical dispersion of each potential for dimensionless position and momentum variables defined in \eref{eqdimensionless}. These dimensionless quantities are bare numbers such that $|X|,|P|  \leq 1$ in a bound potential \cite{Devi}.

At first, eigenvalues ($E_{qm}(n)$) and eigenfunctions ($\psi_n (x)$) of each Schrodinger equation in a given potential are derived and then the related ${\left( \Delta \hat{X} \right)}_{\rm qm}^{2} \: {\left( \Delta \hat{P} \right)}_{\rm qm}^{2}$ is evaluated by averaging on each state in the given potential. Next, the classical dispersion ${\left( \Delta {X} \right)}_{\rm cl}^{2} \: {\left( \Delta {P} \right)}_{\rm cl}^{2}$ is calculated using \eref{eqaverage} with $E_{cl} = E_{qm}$, and compare it with each quantum uncertainty.

Eigenvalues and eigenfunctions can be derived easily by creating a matrix representation for the second derivative operator at first (see \ref{appedixmatrix}); then the related arrays of potential (as a diagonal matrix) are added to the arrays of the second derivative matrix and finally, the eigenvalues and eigenvectors of the summation matrix can be taken using a simple assigned command for this purpose in computer programs. Some useful Matlab codes for educational quantum problems are created in \cite{Garcia}.
 
Quantum uncertainties and classical dispersions for symmetric potentials of \eref{eq4055} are shown in \fref{gasbulbdata} in different quantum states. Also, \fref{sunsets} shows the results of non-symmetric potentials of \eref{eq4056nonsymetric}.

Since $b=2$ in \eref{eq4055} refers to the symmetric harmonic oscillator potential, quantum uncertainty for $b=2$ in \fref{gasbulbdata} is equal to the analytic result of \eref{eqharmonic}, or it is exactly equal to classical dispersion for this potential --independent of the quantum number $n$. Also $b=1$ in \eref{eq4056nonsymetric} refers to bouncing ball potential and again, its quantum uncertainty (see $b=1$ in the \fref{sunsets}) is exactly equal to its classical counterpart, which is in agreement with \eref{eqbouncing}  --independent of $n$. 

In the next two sections, we will argue about $b=1$ in \fref{gasbulbdata}, which belongs to numerical evaluations for symmetric linear potential
\begin{equation}
U(x) = a |x| \qquad -A \leq x \leq A
\end{equation}
with amplitude $A$, and $b=2$ in \fref{sunsets} is related to numerical evaluations for non-symmetric harmonic oscillator  
\begin{equation}
U(x) = \cases{\infty &for $x<0$ , \\
a x^2 &for $x\geq 0$ . \\}
\end{equation}
\section{Comparison between quantum and classical uncertainties for symmetric linear potential}\label{sectionlinear}
Bouncing ball potential is non-symmetric; however, by considering the symmetric potential of $U=a|x|$, we can observe that at least for $n=0$ ($b=1$ in \fref{n_027000}), classical and quantum uncertainties have different values. In fact, bouncing ball potential results in half of the wave functions of symmetric linear potential; in the other words, we will have solutions with $\psi (0)=0$ (odd wave functions). On the other hand, symmetric linear potential inserts all of $\psi (0)=0$ and ${\psi}' (0)=0$ solutions \footnote{Matrix elements of the quantum bouncer \cite{Goodmanson} do not involve even wave functions of symmetric linear potential or those for which ${\psi}' (0)=0$.} and quantum uncertainties in different even $n$ become different from each other quantitatively. Despite this, according to the correspondence principle, in the large $n$, uncertainties become approximately equal for both odd and even solutions.

Considering quantum averages, we seek for averages which cause difference between classical and quantum uncertainties of the linear potential. We know that ${\left\langle {\hat{X}} \right\rangle}$ is zero for the symmetric potentials. Also, it can be derived that ${\left\langle {\hat{P}}\right\rangle}$ and ${\left\langle {\hat{P}}^2 \right\rangle}$ are independent of $n$ (the first moment ${\left\langle {\hat{P}}\right\rangle}$ remains zero since its integral in the position space depends on the value of $\psi (x)$ in bounds and for the second moment ${\left\langle {\hat{P}}^2 \right\rangle}$, we can derive ${\left\langle {\hat{\p}}^2 \right\rangle}= \alpha E(n)$ and ${\left\langle {\hat{P}}^2 \right\rangle} \propto {\left\langle {\hat{\p}}^2 \right\rangle}/E$ from the quantum Virial theorem; then ${\left\langle {\hat{P}}^2 \right\rangle}$ is independent of $n$). Therefore, the difference between classical and quantum uncertainties for $b=1$ in \fref{n_027000} must be created by n-dependence of the second moment of position ${\left\langle {\hat{X}}^2 \right\rangle}$.  
\begin{figure}[h!]
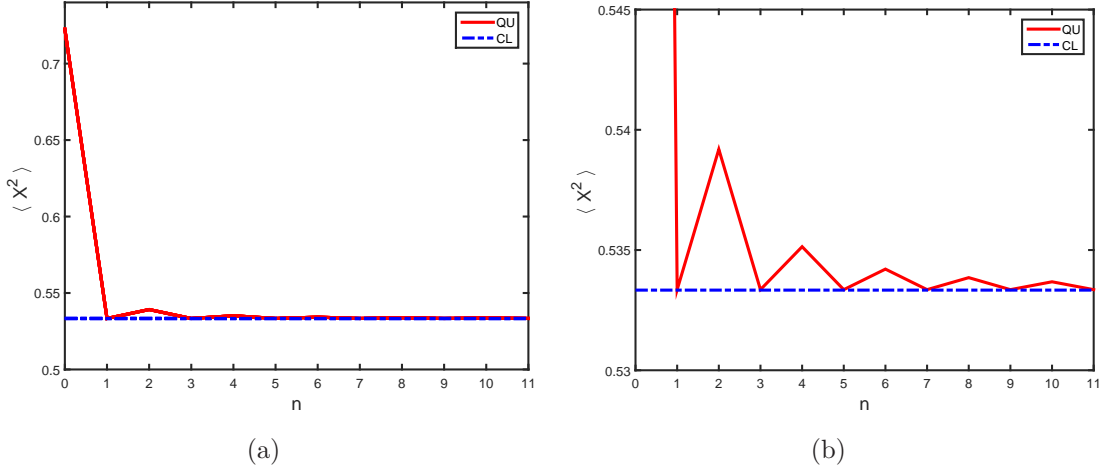

\centering
\subfigure[]{
\includegraphics[width=7.1cm, height=5.5cm]{figure3a}
\label{formal}
}
\subfigure[]{
\includegraphics[width=7.1cm, height=5.5cm]{figure3b}
\label{zoom}
}
\caption{Classical (dashed) and quantum (solid) $\left\langle X^{2} \right\rangle$ belonged to solutions of symmetric linear potential in terms of energy levels $n$ ($X$ is a dimensionless position variable defined in Eq. \eref{eqdimensionless}). \Fref{formal} shows clearly the value for $n=0$, while in \fref{zoom} the same graph is vertically magnified to show other states except $n=0$.}
\label{b_1x2}
\end{figure}
\Fref{b_1x2} displays quantum and classical $\left\langle X^{2} \right\rangle$ for the symmetric linear potential in terms of $n$ and explicitly shows that ${\left\langle {\hat{X}}^2 \right\rangle}_{n=0} \simeq 0.72$ keeps distance from that of other states. We have magnified vertical axis scale for better visibility of other states in \fref{zoom}, such that $0.72$ could not be seen. Difference between quantum uncertainty and classical dispersion at the lowest quantum state of symmetric linear potential ($b=1$ in \fref{n_027000}) arises from the difference between $0.72$ in \fref{formal} with $8/15$, which is related to the classical amount and the amount calculated for wave functions of bouncing ball potential \cite{Goodmanson} (odd wave functions of linear potential). It can be observed that the difference with classical limit for other even states in \fref{b_1x2} is slight and shows decrease. So, this difference is not dramatic in \fref{gasbulbdata} for the next even $n$.
\begin{equation}
\fl \quad {\left\langle {\hat{X}}^2 \right\rangle}_{n=0} > {\left\langle {\hat{X}}^2 \right\rangle}_{n=2} > {\left\langle {\hat{X}}^2 \right\rangle}_{n=4} \cdots \, \simeq {\left\langle {\hat{X}}^2 \right\rangle}_{n=1} = {\left\langle {\hat{X}}^2 \right\rangle}_{n=3} = {\left\langle {\hat{X}}^2 \right\rangle}_{n=5} \cdots 
\end{equation}
and
\begin{equation}
 \lim_{n \to \infty} {\left\langle {\hat{X}}^2 \right\rangle}_{n=2k} = {\left\langle {\hat{X}}^2 \right\rangle}_{n= 2k + 1} = {\left\langle X^{2} \right\rangle}_{\rm cl} \quad , \quad k=0,1,2,... \; .
 \end{equation}
Therefore, uncertainty becomes dependent on $n$.
\section{Comparison between quantum and classical uncertainties for non-symmetric harmonic oscillator}\label{sectionharmonic}
Wave functions are for
\begin{equation}
U(x) = \cases{\infty &for $x<0$, \\
a x^2 &for $x\geq 0$, \\}
\label{eq4056nonsymetricosillator}
\end{equation}
odd solutions of the harmonic oscillator potential located in the interval $[0, \infty )$ with doubled normalization constant. So
\begin{eqnarray}
{\left\langle {\hat{X}}^2 \right\rangle}_{\rm non-symmetric} & = & {\left\langle {\hat{X}}^2 \right\rangle}_{\rm symmetric} = \frac{1}{2} \nonumber ; \\
{\left\langle \; {\hat{P}} \; \right\rangle}_{\rm non-symmetric} & = & {\left\langle\; {\hat{P}} \; \right\rangle}_{\rm symmetric} = \, 0 \nonumber ; \\
{\left\langle {\hat{P}}^2 \right\rangle}_{\rm non-symmetric} & = & {\left\langle {\hat{P}}^2 \right\rangle}_{\rm symmetric} = \frac{1}{2} .
\end{eqnarray}
Non-symmetric ${\left\langle {\hat{X}} \right\rangle}$ can be evaluated analytically, as shown in \ref{appendixmoments}. In the absence of even wave functions, odd wave functions here are starting from $n=0$; therefore, $2n+1$ in \eref{peyvastmean} is transformed to $n$ here.  
\begin{equation}
{\left\langle {\hat{X}} \right\rangle}_{(2n+1) \to n} = \frac{2}{{\pi ^{\frac{1}{2}} }}\frac{{\left( {2n + 1} \right)}}{{\left( {4n + 3} \right)^{\frac{1}{2}} }}\frac{{\left( {2n} \right)!}}{{2^{2n} \left( {n!} \right)^2 }} .
\end{equation}
As a distinctive case, for n=0, we have
\begin{equation}
{\left\langle {\hat{X}} \right\rangle}_{n=0} = \frac{2}{\sqrt{\pi}} \, \frac{1}{\sqrt{3}}
\end{equation}
which results in
\begin{equation}
{\left[ {\left( \Delta \hat{X} \right)}_{\rm qm}^{2} \: {\left( \Delta \hat{P} \right)}_{\rm qm}^{2} \right]}_{n=0} = \left( \frac{1}{2} - \frac{4}{3 \pi } \right) \frac{1}{2} \simeq 0.0378 \; .
\end{equation}
This is in agreement with $b=2$ in \fref{nonsymetric_n_0}. Thus, uncertainty is again dependent on $n$.
\section{Discussion}\label{sectiondisscussion}
In this section, we briefly take a look at two general points which should be considered when studying classical limits.
\subsection{Quantum uncertainty in classical limit}
In addition to topics of sections \ref{sectionlinear} and \ref{sectionharmonic} as the main goals of this article, figures \ref{gasbulbdata} and \ref{sunsets} also show that position-momentum quantum uncertainty relation does not vanish in classical limit. In other words, quantum uncertainty tends to classical evaluations in the classical limit, but not to zero. Namely 
\[ \lim_{n\to\infty} {\left( \Delta \hat{X} \right)}_{\rm qm}^{2} \: {\left( \Delta \hat{P} \right)}_{\rm qm}^{2} = {\left( \Delta {X} \right)}_{\rm cl}^{2} \: {\left( \Delta {P} \right)}_{\rm cl}^{2} \neq 0 \]
Therefore, in the classical limit of $n\to\infty$, quantum particles still have quantum uncertainty in its quantum context. It means that the equality relation between $n\to\infty$ limit of quantum uncertainty and classical dispersion is approaching averages, not approaching concept of quantum uncertainty to that of the classical.
\subsection{Ensemble's interpretation in classical limit}
There are articles such as \cite{Ballentine, Huang} that emphasize that the classical limit of a wave function describes an ensemble of classical particles -not an individual particle. For example, Ballentine et al. \cite{Ballentine}, when challenging Ehrenfest theorem, noted that:
\begin{quote}
\emph{``Generally speaking, the classical limit of a quantum state is not a single classical orbit, but an ensemble of orbits.''}
\end{quote}
Moreover, Huang \cite{Huang}, when using Wigner distribution function \cite{Wigner}, compared with a given probabilistic definition of classical pure ensemble, concluding that:
\begin{quote}
\emph{``A wave function does not describe an individual particle but a classical pure ensemble.''}
\end{quote}

But not all of the classical limits prevent an individual classical particle description. For simplicity, we restrict our study to a one dimensional bound potential of $U(x)$. As an exercise, we can show that in this restricted condition, there are no degenerate wave functions \cite{Griffiths}, so that every energy level $E_n$ belongs to a wave function $\psi_n (x)$.  
Now, we classify these wave functions based on pre and post measured states.

\textbf{1- Before-measurement wave function:}
A wave function before measurement is assumed to be represented by a superposition of all the possible eigenstates.
\begin{equation}
\psi(x) = \sum_{n=0}^{\infty} C_n \; \psi_n (x),
\label{licombination}
\end{equation}
Clearly, $n \to \infty$ limit for such a wave function is meaningless. 
However, we can correspond a $p_i (x)$ to every $|\psi_i (x)|^2$ in terms of distinct energies ($E_{qm} = E_{cl}$). Roughly
\[\fl \left|\psi_i (x) \right|^{2} \longrightarrow p_i (x)\propto \frac{1}{\sqrt{(E_i =E) - U(x)}} ,\qquad p_i (x) \neq p_j (x) \quad \Longleftrightarrow \quad E_i \neq E_j .\]
Then, the set of $\left\lbrace p_0 (x), p_1 (x), p_2 (x),...\right\rbrace$ would describe an ensemble of non-identical particles. These particles would be non-interacting in case it is assumed that independent terms in a linear combination such \eref{licombination} do not interact with each other.

\textbf{2- After-measurement wave function:}
On the other hand, when the measurement is performed, the system is thrown into one of the eigenstates \cite{Sakurai1} of $\mathcal{H}$. In other words
\[\psi (x) \equiv \psi_n (x),\]
then we can use $n \to \infty$ limit for this $\psi (x)$ to correspond it to a specific $p (x)$ (in a given energy $E_{cl} = E_{n}$) and describe the probability density of an individual classical particle or an ensemble of identical particles.

The difference between the classical limit of a wave function in a linear combination of independent eigenfunctions 
\footnote{Special examples for eager student are equation 18 in \cite{Ballentine} and equation 19 in \cite{Huang}; however the second one (free particle) has degeneracy.},
and the classical limit of one of those independent eigenfunctions (here, a non-degenerate energy eigenstate) could be important when encountering various kinds of classical limits through literature. 

\section{Conclusion}
In this note, we narrowly focused on where a quantum feature of ``dependence on the number of states at low energy levels'' seemingly was lost when evaluating the dimensionless uncertainty of bouncing ball and harmonic oscillator potentials, since the quantum results were exactly the same as the classical results. We concluded that this loss was removable with changing potential intervals in terms of parity. In short, this meant including even wave functions in the solutions of the linear potential and excluding them for the quadratic potential by changing potential intervals. Hence it is concluded that the existence of even and odd parity solutions of Schrodinger equation is an inherently quantum mechanical property that is pursuable in the dimensionless averaging of lower energy levels, even in potentials for which Ehrenfest theorem is exact. This could be important, since dimensionless analysis must not eliminate inherently quantum features in general.

Calculations also contain higher powers of position for potential to expand dimensionless analysis to other examples via the numerical method. According to figures \ref{gasbulbdata} and \ref{sunsets}, higher powers of potentials depend on $n$ in both symmetric and non-symmetric forms, while linear and quadratic potentials lose their $n$--dependence in the shape of bouncing ball and harmonic oscillator potentials, respectively. Therefore, we have not paid more attention to higher powers, where n-dependence can be explicitly seen in figures \ref{gasbulbdata} and \ref{sunsets}. Students could study our considerations for even larger powers of $b$, where they eventually might encounter numerical constrains. It would be an instructive challenge to find sources for lack of accuracy (whether or not correlated with execution time) in their computer codes to develop numerical skills.

As other potential examples to study dimensionless uncertainty and compare it with classical dispersion, we can consider supersymmetric partner potentials for the potentials from the power of coordinate ($U(x) \propto x^b$). A review of supersymmetry in quantum mechanics is explained in the article \cite{Khare} or in the book \cite{Kharebook}. In the case of quadratic potential, the supersymmetric partner potentials do not change, except for a constant. Therefore, wave functions $\psi_n (x)$, classical density $p(x)$ and related $x_{max}$ do not change, so that averaging on dimensionless position remains unchanged, unlike averaging on dimensionless momentum that changes according to the changed ${\p}_{max}$, which causes separate levels in the amount of uncertainty (or even classical dispersion) in terms of quantum number $n$, just as separate levels of energy. Along with this special case, supplementary figure data is provided for the uncertainty of supersymmetric partner potentials from powers of coordinate, illustrating this separation and comparing it with \fref{gasbulbdata}. 

Finally, in the discussion section, we mentioned that approaching quantum averaging to that of classical in large quantum numbers does not vanish quantum uncertainty in the classical limit. Also, in the case of non-degenerate states, we have emphasized that it is the classical limit of a ``before-measurement wave function'' that can not describe an individual classical particle, while classical limit of an ``after- measurement wave function'' (a non-degenerate energy eigenstate $\psi_n (x)$) can describe an individual classical particle or an ensemble of identical particles. For a more complete study, we can look forward to extend the classification in terms of every linear combination of independent eigenfunctions (including degenerate states, rather than just a before-measurement wave function) versus one of those independent eigenfunctions.

One pedagogical review of dimensional analysis has been explained in a recent paper \cite{2015} with a variety of examples. Pedagogically speaking, dimensionless analysis, as a secondary branch of dimensional analysis, can be given more attention because of its benefit for analytic and numerical problems. Two helpful examples of using dimensionless analysis are deletion of $\hbar$ in uncertainty and comparison with classical dispersion possible, and development of differential equations as dimensionless ones.

\ack
The author appreciates Dr. K. Aghababaei Samani and Prof. B. Mirza for their scientific supports. Also A. R. Usha Devi and H. S. Karthik should be appreciated for utilizing dimensionless variables which give us a great perspective when dealing with such problems. Finally, I wish to thank R. Garcia, A. Zozulya, and J. Stickney for their MATLAB codes for teaching quantum physics. 

\appendix

\section{Derivation of phase space averages from integrating on position space}\label{averagefequation}
The position probability function is obtained by integrating over the momentum variable $\p$
\begin{equation}
p(x) = \int \rmd \p \, p(x, \p) = \rm{constant}\, . \int \rmd  \p \, \delta \left[ \frac{{\p}^2}{2m} + U(x) - E \right]
\end{equation}
By using the properties $\delta (ax) = \delta (x) / |a|$ and $\delta (x^2 - a^2) = [\delta (x +a ) + \delta (x - a )]/2 |a|$ of the Dirac delta function, the classical probability distribution is reduced to
\begin{eqnarray}
\fl p(x)  =  \rm{constant}\, . \int \rmd  \p 2m\, \delta \left( {\p}^2 + 2m[U(x) - E] \right) = \rm{constant}\, . \sqrt{\frac{2m}{[E -U(x)]}} \nonumber \\
\times \int \rmd  \p \, \left[ \delta \left( {\p} + \sqrt{2m[U(x) - E]} \right) \right. \nonumber \\
\left. + \delta \left( {\p} - \sqrt{2m[U(x) - E]} \right) \right] \propto \frac{1}{\sqrt{E - U(x) }}
\end{eqnarray}
, which is in agreement with \eref{vequation}. In the next step, the phase space averages of any arbitrary function $f(x,\p)$ of position and momentum variables get reduced to those evaluated with the position probability distribution function $p(x)$ as follows:
\begin{eqnarray}
\fl \left\langle f(x,\p) \right\rangle = \int \rmd x \int \rmd \p \; p(x, \p) f(x, \p ) \propto \int \rmd  x \int \rmd  \p \, \delta \left( \frac{{\p}^2}{2m} + U(x) - E  \right) f(x, \p ) \nonumber \\
\fl \propto \int \rmd  x \sqrt{\frac{2m}{E - U(x)}} \int \rmd  \p \left[ \delta \left( {\p} + \sqrt{2m[U(x) - E]} \right) + \delta \left( {\p} - \sqrt{2m[U(x) - E]} \right) \right] f(x, \p ) \nonumber \\
\fl \propto \int \rmd x \sqrt{\frac{2m}{E - U(x)}} \left[ f(x, - \sqrt{2m[E - U(x)]}) + f(x,  \sqrt{2m[E - U(x)]}) \right],
\end{eqnarray}
so that
\begin{eqnarray}
\fl \left\langle f(x,\p )\right\rangle = \frac{1}{2} \int \rmd x\, p(x) \left[ f \left( x , - \sqrt{2m[E -U(x)]} \right) + f \left( x , \sqrt{2m[E -U(x)]} \right)  \right]. 
\end{eqnarray}
 
\section{Matrix representation structure for solving Schrodinger equation}\label{appedixmatrix}
In the numerical solutions, we are looking for eigenvalues and eigenvectors of the Hamiltonian operator
\begin{equation}
\mathcal{H}= - \frac{{\hbar}^2}{2m} \frac{{\rmd}^2}{{\rmd x}^2} + U(x) ,
\end{equation}
in its matrix representation.
Suppose that
$x_{i+1} = x_i + \Delta x$. For a given small
$\Delta x$,
we can write
\begin{equation}
f' (x_i ) \simeq \frac{f(x_{i +1}) - f(x_i)}{\Delta x} \quad , \quad f'' (x_i ) \simeq \frac{f'(x_{i +1}) - f'(x_i)}{\Delta x} .	
\label{eq4057}
\end{equation}
\begin{equation}
\Longrightarrow \qquad f'' (x_i ) \simeq \frac{f(x_{i +2}) -2 f(x_{i +1})  + f(x_i)}{{\Delta x}^2}.
\end{equation}
In Dirac representation,
$\left| {\left. f \right\rangle } \right.$
is a column matrix in the
$x$
basis
\begin{equation}
\left| {\left. f \right\rangle } \right. \rightarrow
\left( \begin{array}{*{20}c}
   f(x_1)  \\
   f(x_2) \\
   \vdots  \\
   f(x_n)  \\
\end{array} \right) .
\end{equation}
For derivative operator $D$, we have
\begin{equation}
D \left| {\left. f \right\rangle } \right. = \left| {\left. f' \right\rangle } \right. ,
\end{equation}
or
\begin{equation}
{\left[ \vphantom{\left( \begin{array}{*{20}c}
   f(x_1)  \\
   f(x_2) \\
   \vdots  \\
   f(x_n)  \\
\end{array} \right)} D \right ]}_{n \times n} 
{\left( \begin{array}{*{20}c}
   f(x_1)  \\
   f(x_2) \\
   \vdots  \\
   f(x_n)  \\
\end{array} \right)}_{n \times 1} = \frac{1}{\Delta x}
{\left( \begin{array}{*{20}c}
 f(x_2) -  f(x_1)  \\
  f(x_3) - f(x_2) \\
   \vdots  \\
   f(x_{n+1}) - f(x_n)  \\
\end{array} \right)}_{n \times 1}.
\end{equation}
Therefore
\begin{equation}
D = \frac{1}{\Delta x} 
\left( {\begin{array}{*{20}c}
   { - 1} & 1 & {} & {} & {} & {}  \\
   {} & { - 1} & 1 & {} & 0 & {}  \\
   {} & {} & { - 1} & 1 & {} & {}  \\
   {} & 0 & {} & { - 1} & 1 & {}  \\
   {} & {} & {} & {} &  \ddots  &  \ddots   \\
\end{array}} \right) .
\end{equation}
In a similar way, for the second derivative, we can write
\begin{equation}
D^{2} = \frac{1}{{\Delta x}^2} 
\left( {\begin{array}{*{20}c}
   1 & { - 2} & 1 & {} & {} & {} & {}  \\
   {} & 1 & { - 2} & 1 & {} & 0 & {}  \\
   {} & {} & 1 & { - 2} & 1 & {} & {}  \\
   {} & 0 & {} & 1 & { - 2} & 1 & {}  \\
   {} & {} & {} & {} &  \ddots  &  \ddots  &  \ddots   \\
\end{array}} \right) .
\end{equation}
According to the definition of a derivative, however, this matrix could be a little different.

Having written potential arrays in the shape of a diagonal matrix, we can have a matrix representation of Hamiltonian. Generally speaking, making a dimensionless Schrodinger equation is more formal at first, but for our purpose, which is averaging on dimensionless quantities of \eref{eqdimensionless}, multipliers like $m$ and $\hbar$ in the Hamiltonian and also, $a$ in the potentials
\eref{eq4055} and \eref{eq4056nonsymetric} 
can be ignored easily.

Note that wave functions naturally tend to zero at $x \to \infty$, but since writing a matrix representation with infinite arrays is impossible, then our wave functions (eigenvectors) do not continue to infinity, so we have to adopt an end for $x$ axis like $\lambda$.
\begin{eqnarray}
&x\in \left[ - \lambda , \lambda \right] \qquad \textbf{\rm{for symmetric potentials}}, \nonumber \\
& x \in \left[ 0 , \lambda \right] \qquad \,\, \textbf{\rm{for non-symmetric potentials}}.
\end{eqnarray}
For a bound state, $\lambda$ can be chosen about a few times more than classical return points by considering enough $x_{i}$ in the interval.
 \section{Moments for half wave functions of symmetric harmonic oscillator}\label{appendixmoments}
Although odd wave functions of the symmetric harmonic oscillator potential are solutions of non-symmetric harmonic oscillator \eref{eq4056nonsymetricosillator}, we evaluate moments for either even or odd wave functions by starting from zero up to infinity, which we call half wave functions, for pedagogical goals or possible implications. The half wave functions of symmetric harmonic oscillator are
\begin{equation}
\fl {\left| \psi_{n} (x) \right|}^{2} = \frac{(2n+1)^{\frac{1}{2}}/{A_n}}{\pi ^{\frac{1}{2}} 2^{n-1} n!}\, \exp \left[ { - \left( {2n + 1} \right)\frac{{x^2 }}{{{A_n}^2 }}{\kern 1pt} } \right]{\kern 1pt} {\kern 1pt} \left\{ {H_n \left[ {\left( {2n + 1} \right)^{\frac{1}{2}} \frac{x}{{A_n}}} \right]} \right\}^2 
\end{equation}
, where ${A_n}$ satisfies $( n + 1/2 ) \hbar \omega = 1/2 m {\omega}^2 {{A_n}^2 } $. Note that wave functions are normalized in the interval $[0,\infty )$.
The moments are as follows
\begin{equation}
\fl {\left\langle x^k \right\rangle}_{n} = {A_n^k} \frac{(2n+1)^{- \frac{k}{2}}}{\pi ^{\frac{1}{2}} 2^{n-1} n!}\, \int\limits_{0}^{\infty} y^k \exp \left( -y^2 \right) \, {\left[ {H_n} (y) \right]}^2 \rmd y , \qquad y=\frac{(2n+1)^{\frac{1}{2}}}{{A_n}} x
\end{equation}
or
\begin{equation}
{\left\langle X^k \right\rangle}_n = \frac{(2n+1)^{- \frac{k}{2}}}{\pi ^{\frac{1}{2}} 2^{n-1} n!}\, \int\limits_{0}^{\infty} y^k \exp \left( -y^2 \right) \, {\left[ {H_n} (y) \right]}^2 \rmd y .
\label{eq3059}
\end{equation}
There is a useful tabled integral as \cite{Spiegel}
\begin{equation}
\mathop \int \limits_0^\infty  x^m \exp ( - \alpha x^2 ) \rmd x = \frac{{{\rm \Gamma }\left[ {(m + 1)/2} \right]}}{{2\alpha ^{\frac{{m + 1}}{2}} }}
\end{equation}
, where $\Gamma (x)$ indicates the Gamma function. The moments can be written by a
summation of these integrals. In this way, mean values can be obtained. However, it may not simply give us a general term in respect to $n$ for \eref{eq3059}. Instead, it can be seen that for even numbers $k$, integrand is an even
function. Hence
\begin{equation}
\fl {\left\langle X^{2m} \right\rangle}_n  = \frac{{(2n + 1)^{ - m} }}{{\pi ^{\frac{1}{2}} 2^n n!}}\mathop \int \limits_{ - \infty }^\infty  y^{2m} \exp ( - y^2 ) \left[ {H_n \left( y \right)} \right]^2 \rmd y ,\qquad k=2m,\; m\in \mathbb{N}
\end{equation}
They can be evaluated using orthogonality and recurrence relations for Hermite functions (see, for example, exercise $13.1.11$ of \cite{Arfken}) and substituting recurrence relation again for even levels higher than $m=1$. For instance,
\begin{equation}
{\left\langle X^4 \right\rangle} _n  = \frac{3}{2} \, \frac{{n^2  + n + \,1/2}}{{\left( {2n + 1} \right) }^2} , \qquad \lim_{n\to \infty} {\left\langle X^4 \right\rangle}_n = \frac{3}{8} .
\label{peyvastclassicalfourtmom}
\end{equation}
Here we return to the first moment by using derivative formulas \cite{Spiegel}
\begin{equation}
\fl \frac{\rmd}{{\rmd x}}\left[ {\exp ( - x^2 ) H_n \left( x \right)} \right] =  - \exp ( - x^2 ) H_{n + 1} \left( x \right) , \qquad \frac{\rmd}{{\rmd x}}H_n \left( x \right) = 2nH_{n - 1} \left( x \right)
\end{equation}
and integral recurrence relation
\begin{eqnarray}
\fl \mathop \int \limits_0^\infty  x \exp ( - x^2 ) \left[ {H_n \left( x \right)} \right]^2 \rmd x = \frac{1}{2}\mathop \int \limits_0^\infty  \exp ( - x^2 ) H_{n + 1} \left( x \right)H_n \left( x \right)\rmd x \nonumber \\
+ n\mathop \int \limits_0^\infty  \exp ( - x^2 ) H_n \left( x \right)H_{n - 1} \left( x \right)\rmd x .
\end{eqnarray}
Applying derivative formulas and then integrating by parts, we have
\begin{eqnarray}
\fl \mathop \int \limits_0^\infty  x \exp ( - x^2 ) \left[ {H_n \left( x \right)} \right]^2 \rmd x = \frac{1}{2}\left[ {H_n \left( 0 \right)} \right]^2  + n!\mathop \sum \limits_{k = 1}^n \frac{{2^k }}{{\left( {n - k} \right)!}}\left[ {H_{\left( {n - k} \right)} \left( 0 \right)} \right]^2 \nonumber \\
=  - \frac{1}{2}\left[ {H_n \left( 0 \right)} \right]^2  + n!\mathop \sum \limits_{k = 0}^n \frac{{2^k }}{{\left( {n - k} \right)!}}\left[ {H_{\left( {n - k} \right)} \left( 0 \right)} \right]^2 .
\end{eqnarray}
Remembering
\begin{equation}
H_{2n} \left( 0 \right) = \left( { - 1} \right)^n \frac{{\left( {2n} \right)!}}{{n!}},\quad H_{2n + 1} \left( 0 \right) = 0
\end{equation}
and
\begin{equation}
\mathop \sum \limits_{k = 0}^n \frac{{\left[ {2\left( {n - k} \right)} \right]!}}{{\left[ {\left( {n - k} \right)!} \right]^2 }}2^{2k}  = \frac{{\left( {2n + 1} \right)!}}{{\left( {n!} \right)^2 }} ,
\label{firstmomentseri}
\end{equation}
we can derive
\begin{eqnarray}
\fl \mathop \int \limits_0^\infty  x \exp ( - x^2 ) \left[ {H_{2n} \left( x \right)} \right]^2 \rmd x =  - \frac{1}{2}\left[ {\frac{{\left( {2n} \right)!}}{{n!}}} \right]^2  + \left( {2n} \right)!\mathop \sum \limits_{k = 0}^n \frac{{\left[ {2\left( {n - k} \right)} \right]!}}{{\left[ {\left( {n - k} \right)!} \right]^2 }}2^{2k} \nonumber \\
=  - \frac{1}{2}\left[ {\frac{{(2n)!}}{{n!}}} \right]^2  + \left( {2n + 1} \right)\left[ {\frac{{\left( {2n} \right)!}}{{n!}}} \right]^2  = \left( {2n + \frac{1}{2}} \right)\left[ {H_{2n} \left( 0 \right)} \right]^2
\end{eqnarray}
and
\begin{eqnarray}
\fl \mathop \int \limits_0^\infty  x \exp ( - x^2 ) \left[ {H_{2n+1} \left( x \right)} \right]^2 \rmd x & = \left( {2n+1} \right)!\mathop \sum \limits_{k = 0}^n \frac{{\left[ {2\left( {n - k} \right)} \right]!}}{{\left[ {\left( {n - k} \right)!} \right]^2 }}2^{2k+1} \nonumber \\
& = 2 \left( {2n+1} \right)!\mathop \sum \limits_{k = 0}^n \frac{{\left[ {2\left( {n - k} \right)} \right]!}}{{\left[ {\left( {n - k} \right)!} \right]^2 }}2^{2k} \nonumber \\
& = 2\left[ {\frac{{\left( {2n + 1} \right)!}}{{n!}}} \right]^2  = 2\left( {2n + 1} \right)^2 \left[ {H_{2n} \left( 0 \right)} \right]^2 .
\end{eqnarray}
So, the first moment leads to
\begin{eqnarray}
{\left\langle X \right\rangle}_{2n} & = & \frac{2}{{\pi ^{\frac{1}{2}} }}\frac{{\left( {2n + 1} \right)}}{{\left( {4n + 1} \right)^{\frac{1}{2}} }}\frac{{\left( {2n} \right)!}}{{2^{2n} \left( {n!} \right)^2 }} ; \\
{\left\langle X \right\rangle}_{2n + 1} & = & \frac{2}{{\pi ^{\frac{1}{2}} }}\frac{{\left( {2n + 1} \right)}}{{\left( {4n + 3} \right)^{\frac{1}{2}} }}\frac{{\left( {2n} \right)!}}{{2^{2n} \left( {n!} \right)^2 }} .
\label{peyvastmean}
\end{eqnarray}
The limit of $n \to \infty$ can be evaluated by using \footnote{Equations~(\ref{firstmomentseri})~and~(\ref{cllimitoffirstmoment}) have been obtained by \textit{MAPLE} mathematical software.}
\begin{equation}
\mathop {{\rm lim}}\limits_{n \to \infty } \frac{{n^{\frac{1}{2}} \left( {2n} \right)!}}{{2^{2n} (n!)^2 }} = \frac{1}{{\sqrt \pi  }}, \label{cllimitoffirstmoment}
\end{equation}
thus
\begin{equation}
\mathop {{\rm lim}}\limits_{n \to \infty } {\left\langle X \right\rangle}_{2n}  = \mathop {{\rm lim}}\limits_{n \to \infty } {\left\langle X \right\rangle}_{2n + 1}  = \frac{2}{{\pi ^{\frac{1}{2}} }} \, \mathop {{\rm lim}}\limits_{n \to \infty } \, \frac{{n^{\frac{1}{2}} \left( {2n} \right)!}}{{2^{2n} \left( {n!} \right)^2 }} = \frac{2}{\pi }.
\label{peyvastclassicalfimom}
\end{equation}
Equations~(\ref{peyvastclassicalfourtmom})~and~(\ref{peyvastclassicalfimom}) can be checked against the classical results when we average over a quadratic classical density function.

\newpage
\textbf{Supplementary Figure}
\renewcommand{\figurename}{Caption}
\begin{figure}[h!]
\centering
\includegraphics[width=15cm, height=6.5cm]{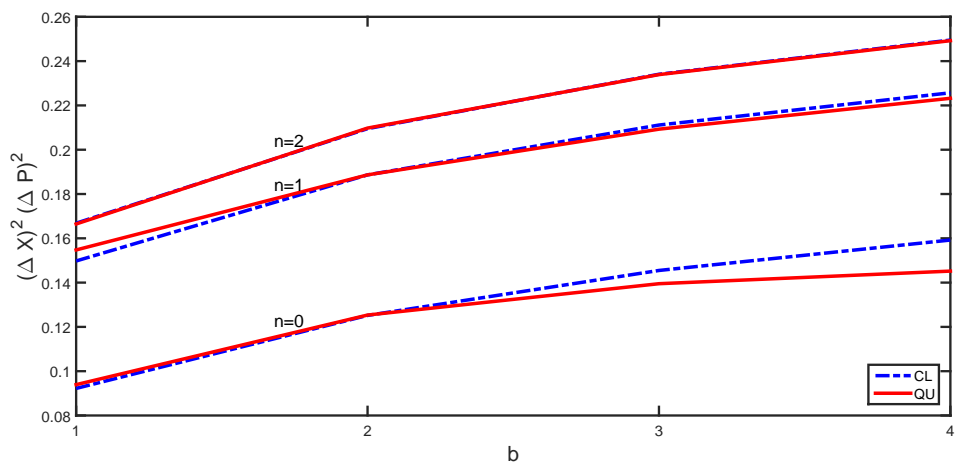}
\caption{Classical (dashed) and quantum (solid) uncertainty of dimensionless variables for supersymmetric  partner potentials of equation \eref{eq4055}($b=1,2,3,4$) on three first levels}
\label{Supplementary}
\end{figure}

\end{document}